\documentclass{aa}
\usepackage{graphicx}
\begin{document}
   \title{Black hole mass and binary model for BL Lac 
		object OJ 287}

   \author{F.K. Liu\inst{1,2}
          \and
          Xue-Bing Wu\inst{2}
          }

   \offprints{F.K. Liu}

   \institute{Department of Astronomy and Astrophysics, Gothenburg
		University \& Chalmers University of Technology,
		41296 Gothenburg, Sweden 
         \and
             Department of Astronomy, Peking University, 100871 Beijing, 
		China \\
             \email{fkliu@bac.pku.edu.cn, wuxb@bac.pku.edu.cn}
             }

   \date{Received ***; accepted ***}

   \abstract{
	Recent intensive observations of the BL Lac object OJ 287 raise
	a lot of questions on the models of binary black holes, processing
	jets, rotating helical jets and thermal instability of slim 
	accretion disks. After carefully analyzing their 
	radio flux and polarization data, Valtaoja et al. (\cite{valtaoja00})
	propose a new binary model. Based on the black 
	hole mass of $4 \times 10^8 {\rm M_\odot}$ estimated with the tight
	correlations of the black hole masses and the bulge luminosity 
	or central velocity dispersion of host galaxies, we computed the 
	physical parameters of the new binary scenario. The impact of the 
	secondary on the accretion disk around the primary black hole
	causes strong shocks propagating inwards and outwards, whose
	arrival at the jet roots is identified with the rapid increase of 
	optical polarization and the large change of polarization angle 
	at about 0.30 yr after the first main optical flare. An increase of 
	optical polarization, a large rotation of positional angle 
	and a small synchrotron flare at 2007.05 between the optical
	outbursts at 2006.75 and 2007.89 are expected by the model. 
	With the estimated parameters, we predicated an increase 
	of $\gamma$-ray flux appearing about 5 days after the first
	optical/IR peak, which is consistent with the EGRET observations.

   }

\titlerunning{Binary model for OJ 287}

   \maketitle
%

\section{Introduction}

OJ 287 (0851+202) at red-shift $z = 0.306$ is one of the most active and
best studied BL Lac objects. While it was identified in 1967 (Dickel et al. 
\cite{dickel67}), its optical observations were dated back to 
1890s, which makes it be one of the best candidates for searching for
long-term periodicity. Based on the historical optical light curves, 
Sillanp\"a\"a et al. (\cite{sillanpaa88}) found the large optical 
outbursts of OJ287 recurrent with a period of 11.65 yr and proposed a 
binary black hole model, which predicted an outburst in the 1994 fall. 

Although the predicted large optical outbursts were detected in 1994
and 1995, it was found that the outbursts
are double-peaked and the binary black hole model 
has to be modified (Sillanp\"a\"a et al. 
\cite{sillanpaa96}). Among the proposed models to explain the 
periodicity of OJ 287, those of binary black holes
(Lehto \& Valtonen \cite{lehto96}), procession jet (Katz 
\cite{katz97}), rotating helical jets (Villata et al. \cite{villata98}) and
the thermal instability of slim accretion disk (Liu et al. \cite{liu95})
can explain the periodicity in optical band well. However, radio and 
polarization observations
show that the first of the two peaks is narrow and thermal, and the
second one is broad and synchrotronic (Valtaoja et al. \cite{valtaoja00};
Pursimo et al. \cite{pursimo00}). Both flares should be synchrotronic and
identical in the processing/helical jet models while thermal and similar 
to each other in the binary black hole model (Valtaoja et al. 
\cite{valtaoja00}). Another difficulty with the binary model is that it 
requires an extremely massive primary of $M= 1.5\times 10^{10} {\rm M_\odot}$ 
(Pietil\"a \cite{pietila98}), which is at least one order of magnitude 
larger than the super-massive black hole masses of other BL Lac objects 
(Barth et al. \cite{barth02}; Wu et al. \cite{wu02}; Falomo et al. 
\cite{falomo02}).

In the binary black hole model, the outburst thermal optical/IR photons
should be up-scattered into $\gamma$-ray regime by the relativistic 
electrons in the jets via inverse Compton scattering process. From the
parameters given by Pietil\"a (\cite{pietila98}), the estimated time for
photons to travel from the impact site to the jets is about 20 days and 
implies a delay of about three weeks between the optical/IR 
outbursts and high energy $\gamma$-ray radiations. Such a large lag
is not consistent with the EGRET $\gamma$-ray observations (Shrader et 
al. \cite{shrader96}).

To resolve the difficulties of the models mentioned above, Valtaoja et al.
(\cite{valtaoja00}) suggest a new binary scenario to explain the observations
of radio flux and polarization. In their model, the first thermal flare of the 
outbursts is caused when a secondary black hole hyper-sonically passes 
through the accretion disk around the primary, and the second synchrotron
flare after a viscous time scale $t_{\rm vis}$ is from the jet as the 
perigee passage enhances the accretion rate in the accretion disk,
leading to increased jet mass flux and formation of strong shocks down 
the jet. As for such a scenario a constant period agrees with
all the sharp outbursts (the first peak), a precession of the binary orbit 
and an extreme primary black hole are not needed (Valtaoja et al. 
\cite{valtaoja00}). However, as the lack of accurate knowledge of the 
primary black hole mass, Valtaoja et al. (\cite{valtaoja00}) did not estimate 
the parameters of the new system. In this letter, we extend 
Valtaoja et al's model to explain all the observations and try to estimate
the physical parameters of the new scenario. 

After a brief summary of the observations in
Sec.~\ref{sec:obs}, we describe the model in Sec.~\ref{sec:model}. 
In Sec.~\ref{sec:detail}, we try to estimate the parameters of the model.
Our discussions and conclusions are presented in Sec.~\ref{sec:dis}. 
$H_0 = 50 \, {\rm km\, s^{-1}\, Mpc^{-1}}$ and $q_0 = 0.5$ are used 
throughout the paper.


\section{Observations}
\label{sec:obs}

The optical outbursts of OJ287 have a period of $P = 11.886 \, {\rm yr}$ 
(Kidger \cite{kidger00}) and are double-peaked with a 
spacing of 416 days (Sillanp\"a\"a et al. \cite{sillanpaa96}; Kidger 
\cite{kidger00}). The first flare is thermal and narrow without corresponding 
radio outburst and the second one is non-thermal and broad with strong 
outburst in radio (Valtaoja et al. \cite{valtaoja00}; Pursimo et al. 
\cite{pursimo00} and references therein).  The variability in the IR 
band is similar to that in the optical band (Kidger et al. \cite{kidger95};
Pursimo et al. \cite{pursimo00}). The first flare in the nearest 1994-1995 
outbursts peaked at JD1994.86 and the second one did at 1996.00. A small 
UV flare was detected during the first optical flare and no observation 
was made around the second one (Pursimo et al. \cite{pursimo00}). 

The observations made around JD1994.88 showed no evidence for any
X-ray outburst, but weak flux variations of time-scale as short as 
the time limit of the instrument (96 minutes) were detected (Idesawa et 
al. \cite{idesawa97}; Pursimo et al. \cite{pursimo00}). EGRET observations
integrating from 10 to 15 November 1994 showed that in the $\gamma$-ray 
band the 
object was three times brighter than earlier, when it was in a low 
optical state (Shrader et al. \cite{shrader96}; Pursimo et al. 
\cite{pursimo00})

The optical polarization is very low during the first flare but very high
during the second flare. A substantial increase of optical polarization 
and a large change of positional angle within one day at JD1995.16 
have been observed (Smith et al. \cite{smith87}; Takalo 
\cite{takalo94}; Pursimo et al. \cite{pursimo00}; Efimov et al. 
\cite{efimov02}). The variability of radio polarization is similar but 
with a time delay to that of optical polarization (Valtaoja et al. 
\cite{valtaoja00}; Pursimo et al. \cite{pursimo00}). 

The host galaxy of OJ 287 has been marginally resolved recently and 
has an absolute magnitude of $M_{\rm R} = - 23.23$ (Heidt et al. 
\cite{heidt99})


\section{A revised binary black hole model}
\label{sec:model}

The new binary scenario initially proposed by Valtaoja et al. 
(\cite{valtaoja00}) and described in the introduction can be extended
to explain the optical polarization, the X-ray variation and the $\gamma$-ray
radiation of OJ287. 

In addition to an enhancement of the accretion rate, the perigee 
passage also causes strong shocks in the accretion disk, 
which propagate inwards and outwards at the speed of sound. The shocks
arrive at the jet roots after a sound traveling time scale $\Delta t_{\rm s}$
and cause cause a very small
increase of the jet flow and shocks down the jets, compressing
the magnetic fields in the jets and therefore causing a rapid 
increase of the optical polarization and large rotation of 
positional angle about $\Delta t_{\rm s} \simeq 0.30 \, {\rm yr}$ after the first
optical flare. 

The thermal optical/IR photons of the first main flare feed the 
jets of the primary black hole and are up-scattered into the 
$\gamma$-ray regime by the relativistic electrons in the jets via inverse 
Compton scattering. The $\gamma$-ray outbursts, which have a
delay of light traveling time from the impact site to the jets, may 
have been detected by EGRET. The detected weak X-ray variability 
may be generated around the secondary
black hole, as  the secondary is also an X-ray radiation source.

The apogee passage of the secondary black hole through the accretion disk
may be too weak and undetectable.


\section{Parameters of the model}
\label{sec:detail}

\subsection{Masses of the black holes}
\label{sec:mass}

It was found recently that the masses of the center black holes in
active and inactive galaxies follow tight correlations with the bulge 
velocity 
dispersions (e.g. Gebhardt et al. \cite{gebhardt00}; Merritt \& 
Ferrarese \cite{merritt01a,merritt01b}) and the bulge luminosities 
(e.g. Magorrian et al. \cite{magorrian98}; McLure \& Dunlop 
\cite{mclure02}). The host galaxy of OJ287 was marginally resolved only 
recently of an effective radius $r_{\rm e} = 0.72''$ and an 
absolute magnitude $M_{\rm R} = -23.23$ (Heidt et al. \cite{heidt99}). 
A black hole mass of $M \simeq 4.6 \times 10^8 {\rm M_\odot}$ is obtained
with the absolute magnitude together with the correlation of black hole mass 
and bulge luminosity ($M - L_{\rm bulge}$) for active galaxies given by 
McLure \& Dunlop (\cite{mclure02}). With the morphological data  of the
host galaxy in 
Heidt et al. (\cite{heidt99}), we estimated the central velocity 
dispersion $\sigma$, using the 
fundamental plane of a tight correlation of velocity dispersion, 
effective radius and average surface brightness for elliptical galaxies
(Bettoni et al. \cite{bettoni01}). A second value of the black hole
mass $M \simeq 3.2 \times 10^8 {\rm M_\odot}$ was derived with the estimated 
velocity dispersion $\sigma$ and the $M - \sigma$ correlation given by 
Merritt \& Ferrarese (\cite{merritt01b}). The two estimated values of 
the black hole mass agree well with each other and are consistent with 
the upper limit $M \la 10^9 {\rm M_\odot}$ obtained by Valtaoja et al. 
(\cite{valtaoja00}). We used their mean value $M = 4\times 10^8 {\rm 
M_\odot}$ in the following calculation.

For the secondary, one can set only an upper limit $m / M_\odot \la 
10^8$, if the variability of the X-ray flux is from the accreting gas 
around it.


\subsection{Standard thin accretion disk or ADAF?}
\label{sec:disk}

While it has been suggested that the accretion disk in BL Lac objects
is optically thin (Rees et al. \cite{rees82}; Cavaliere \& D'Elia 
\cite{cavaliere02}), it is still possible that OJ 287 is exceptional
and has an optically thick and geometrically thin accretion disk (e.g. 
Sundelius et al. \cite{sundelius97}; Pietil\"a et al. \cite{pietila99}).

For an accretion disk, the viscous time-scale is $t_{\rm vis} \simeq
R/v \simeq R^2 /\nu = R t_{\rm s} /(\alpha H) $ where $t_{\rm s} = R / 
c_{\rm s}$.
If we take $t_{\rm vis} \simeq 1.14/(1 + z)\, {\rm yr}$ and $t_{\rm s} 
\simeq 0.30 /(1 + z) \, {\rm yr}$, we have $H /R \ga 0.3$ for 
$\alpha \la 1$, which implies a geometrically thick disk. If we assume 
that large rotation of optical polarization angle (PA) around the time of the 
first optical outburst is caused by the passage of the secondary through
the accretion disk, the nearly simultaneity between the PA rotation and 
optical outburst implies that the outburst radiation region becomes 
optically thin immediately when it is generated by the passage.
So, it is naturally to assume the accretion disk being optically thin 
and geometrically thick, i.e. an advection-dominated accretion flow (ADAF). 

For an ADAF, the self-similar solutions of sound speed $c_{\rm s}$ and 
radial velocity $v$ are (Narayan et al. \cite{narayan00}), respectively,
$c_{\rm s}   =  c_0 v_{\rm K}$ and $v  =  v_0 v_{\rm K}$, where 
$v_{\rm K}$ is the Keplerian velocity, $c_0 = (2/5)^{1/2}$ and 
$v_0 = 3 \alpha /5$ for the politropic index $\gamma = 5/3$. 
Therefore, we have 
\begin{equation}
 \alpha  =   {\sqrt{10} \over 3} {t_{\rm s} \over t_{\rm vis}} = 0.28 ,
\end{equation}
which is very close to the favored value $\alpha = 0.25$ in 
literature (Narayan et al. \cite{narayan98}). 
The value of $\alpha$ slightly depends on the polytropic index 
$\gamma$ and we have $\alpha \simeq 0.31$ for $\gamma = 7/5$. 


\subsection{Parameters of the binary system}
\label{sec:para}

If the secondary black hole is comparable to the primary, the primary 
would rotate around their mass center with a large amplitude, which leads 
up to in a significant sinusoidal twist of jets. Such a twist was not 
detected by VLBI (e.g. Tateyama et al. \cite{tateyama96}; Roberts et al. 
\cite{roberts87}). So, we have $m \ll M$ and the major axis $a$ of 
the secondary orbit is
\begin{equation}
a \simeq {1 \over 2} \left({P c \over \pi r_{\rm G} \left(1+
	z\right)}\right)^{2/3} r_{\rm G} 
	= 4.1\times 10^2 M_8^{-2/3} r_{\rm G} ,
\end{equation}
where $r_{\rm G}$ is the Schwarzschild radius of the primary 
and $M_8 = M / \left(4 \times 10^8 {\rm M_\odot}\right)$.

The pericenter passage of the orbit is at radius 
\begin{equation}
R_{\rm min} \simeq c_{\rm s} t_{\rm s} /(1 + z) = 88 M_8^{-2/3} r_{\rm G} .
\end{equation}
If the accretion disk is a standard thin high-$\alpha$ disk, the orbit 
of the secondary may be coplanar with the accretion disk due to 
disk-secondary black hole interaction (Ivanov et al. 
\cite{ivanov98}; cf. Valtaoja et al. \cite{valtaoja00}). However, the 
situation is unclear for an advection-dominated accretion flow, as the 
interaction between the secondary and the disk is negligible (Narayan 
\cite{narayan00a}) and the coplanar time scale is of order of viscous 
time scale or larger
(Ivanov et al. \cite{ivanov98}; Scheuer \& Feiler \cite{scheuer96}). In the 
following discussion, we assumed that the orbit of the secondary is 
roughly coplanar with the accretion disk and the secondary passes 
through the accretion disc twice per orbit. Then, the 
eccentricity of the orbit is $e \simeq 0.78$ and the apogee is at radius
\begin{equation}
	R_{\rm max} = 7.2\times 10^2 M_8^{-2/3} r_{\rm G} .
\end{equation}

The IR/optical photons, which are up-scattered into $\gamma$-ray regime by 
the relativistic electrons in the jets, travel from the pericenter passage 
to the jets within a time scale 
$\Delta t_\gamma \simeq R_{\rm min} / c = 4.0 M_8$ days. So, the model predicts
an increase of $\gamma$-ray flux with a lag of $(1+z) \Delta t_\gamma \simeq
5$ days after the main flare, which is consistent with the EGRET 
observations (Shrader et al. \cite{shrader96}; Pursimo et al. 
\cite{pursimo00}).

\subsection{The apogee passage}
\label{sec:apogee}

From the self-similar solutions (Narayan et al. \cite{narayan00}), the 
ratio of the mass densities between the two passages is 
$\rho_{\rm max} / \rho_{\rm min} \simeq (R_{\rm min} /R_{\rm max})^{3/2}= 
4.3 \times 10^{-2}$. As the radiation of optically thin hot gas is 
bremsstrahlung 
and proportional to the density square, the radiation is at least three 
order of magnitudes fainter at the apogee than at the pericenter. The 
difference of temperature, $T_{\rm max}/ T_{\rm min} \sim (R_{\rm min} / 
R_{\rm max}) = 1.2\times 10^{-1}$, makes the radiation at $R_{\rm max}$ 
even weaker and in lower frequency. Therefore, the burst at the apogee 
is undetectable.

%

\section{Discussions and Conclusions}
\label{sec:dis}

Basing on the observations of OJ 287 in multi-frequencies and of the host
galaxy, we extended and estimated the parameters of the revised binary 
model initially
suggested by Valtaoja et al. (\cite{valtaoja00}). In the model, the mass
of the primary black hole is $4\times 10^8 {\rm M_\odot}$, which is consistent
with the upper limit obtained by Valtaoja et al. (\cite{valtaoja00}) and
about 40 times smaller than what the L\&V's model has given (Pietil\"a
\cite{pietila98}). As the semi-major axis is, respectively, about 
$30 r_{\rm G}$ in the L\&V's model and about $4.1 \times 
10^2 r_{\rm G}$ in the revised model, the two black holes are about ten times
closer in the L\&V's model than in ours (in units of the Schwarzschild
radius). With such a less massive primary black hole and larger semi-major 
axis, the gravitational radiation is much weaker in the new scenario,
though we cannot exactly estimate its value as we have only an upper 
limit of the companion mass. It can be expected that the lifetime of the 
binary system is much longer in the new model than that in the L\&V's 
which is only about $10^4 \, {\rm yr}$. 

As we took the constant period of $P = 11.886$ yr as in Kidger 
(\cite{kidger00}), the predicted next outbursts (Kidger 
\cite{kidger00}) are that the first thermal optical flare is 
at 2006.75 (1 Oct 2006) and the second synchrotron flare is at about 
2007.89 (21 Nov 2007). The predictions are consistent within the 
uncertainties of the period  with those given
by Valtaoja et al. (\cite{valtaoja00}). In addition, 
an increase of optical polarization and large rotation of PA at 2007.05 (18 
Jan, 2007) and brightening in $\gamma$-ray on 5 Oct, 2006 (at 2006.76) are 
expected by our model. 

We identify the small X-ray variations observed about one week after the
first flare with radiations from the hot gas around the secondary black hole. 
However, it is possible that the X-ray variability is not relevant to 
the pericenter passage event and cannot be detected during the 2006-2007
outbursts. Another possibility is that the arrival signal of sound wave 
traveling from the impact site to the jets is the
variation in X-ray continuum instead of optical polarization. 
If so, an extremely high eccentricity $e = 0.98$ is needed. The 
orbit of the secondary in such a system is completely embedded in the thick 
accretion disk and no outburst can be observed. Any binary system with 
such an extremely-high initial eccentricity would very rapidly reduce its 
eccentricity to $e \la 0.8$ due to gravitational radiation and the 
interaction between the secondary and the accretion disk (cf Artymowicz 
\cite{artymowicz92}).

If the traveling time scale of shock waves is as short as $t_{\rm s} /2$, the 
eccentricity of the binary system is $e \simeq 0.87$ and most 
part of the orbit must be embedded in the accretion disk. No sharp 
optical outbursts can be detected. If the traveling time scale of the 
shock waves is longer than $t_{\rm s}$, e.g. $2 t_{\rm s}$, the orbit becomes very 
circular with $e = 0.66$ and the two passages are very close to each 
other and $R_{\rm min}/R_{\rm max} \simeq 0.21$. The outbursts caused by the 
apogee passage become significant and detectable. In this case, an 
intersection of two periodic variabilities of $p = 11.886\, {\rm yr}$ should
have been observed. One possibility is that the orbit is circular 
of $a \simeq 640 r_{\rm G}$ (Note that the observed period $P_{\rm obs} = 
P_{\rm orbit} /2$) and we have actually observed two equal passages with
$R_{\rm min} \simeq R_{\rm max} \simeq a$. In this case, the traveling time 
scale of shock waves is about $7.3 t_{\rm s} \simeq 2.2 \, {\rm yr}$, which
implies an unreasonably large $\alpha \simeq 2.0$. Our conclusion is that
the shock traveling time scale of shock waves is between about $t_{\rm s}/2$
and $2 t_{\rm s}$. It is reasonable to assume that the rapid increase of the 
optical polarization and the large rotation of positional angle after about 
0.30 yr following the first flare are caused by the shock waves arriving
at the jet foot-points. As the shock waves may increase the mass flow into
jet and cause shocks down the jets, a small synchrotron outburst 
corresponding to the change of optical polarization is also expected. 

We have assumed that the orbit of the secondary is roughly coplanar 
with the accretion disk in the calculation of the eccentricity $e$ and 
$R_{\rm max}$. If they are not roughly coplanar with each other, the system
would have a much higher eccentricity and closer apogee. This  
problem has already been discussed before. 

We used the self-similar solutions to a stationary and vertically
averaged one-dimension ADAF. Two-dimensional 
numerical simulation (Igumenshchev \& Abramowicz \cite{igo00}) showed
that an ADAF with higher $\alpha$ ($\ga 1$) is convectively unstable.
For an ADAF with lower viscosity ($\alpha \la 0.1$), no bipolar or large
scale circular motion is present and $t_{\rm vis}$ should be 
passage-independent. For ADAF with a moderate viscosity 
$\alpha \sim 0.3$, while a self-similar solution is adequate, 
time-dependent bipolar outflows and large scale circulations are 
important (Igumenshchev \& Abramowicz \cite{igo00}). Thus, a 
passage-dependent time scale $t_{\rm vis}$ is not inconsistent with 
our model. However, to fully understand the time dependences of 
the inner accretion disk, two or even three 
dimensional numerical simulations are needed. In our model, a
time-dependent $t_{\rm vis}$ gives  a time-dependent viscous
parameter $\alpha$. Other estimated parameters are independent of
$t_{\rm vis}$.

\begin{acknowledgements}
        We are grateful to the anonymous referee for his/her helpful 
	comments.
	FKL acknowledges support by the Swedish Natural Sciences
	Research Council (NFR). He thanks the director and staff of the 
	Department of Astronomy and Astrophysics, G\"oteborg University 
	\& Chalmers University of Technology, for their warm hospitality 
	during his stay.
\end{acknowledgements}


\end{document}